\documentclass[
]{ceurart}

\usepackage{listings}
\usepackage{CJKutf8}
\usepackage{url}

\begin{document}


\conference{21st Conference on Information and Research Science Connecting to Digital and Library Science, February 20-21 2025, Udine, Italy}

\title{Recent Developments in Deep Learning-based Author Name Disambiguation}

\author[1]{Francesca Cappelli}[%
email=francesca.cappelli9@unibo.it,
url=https://www.unibo.it/sitoweb/francesca.cappelli9,
]
\cormark[1]

\author[1,2,3]{Giovanni Colavizza}[%
orcid=https://orcid.org/0000-0002-9806-084X,
email=giovanni.colavizza@unibo.it,
url=https://github.com/Giovanni1085,
]

\author[1,3,4]{Silvio Peroni}[%
orcid=https://orcid.org/0000-0003-0530-4305,
email=silvio.peroni@unibo.it,
url=https://github.com/essepuntato,
]

\address[1]{Department of Classical Philology and Italian Studies, University of Bologna, Bologna, Italy.}
\address[2]{Department of Communication, University of Copenhagen, Copenhagen, Denmark.}
\address[3]{Digital Humanities Advanced Research Center (DHARC), University of Bologna, Bologna, Italy.}
\address[4]{Research Centre for Open Scholarly Metadata, University of Bologna, Bologna, Italy.}

\cortext[1]{Corresponding author.}

\begin{abstract}
Author Name Disambiguation (AND) is a critical task for digital libraries aiming to link existing authors with their respective publications. Due to the lack of persistent identifiers used by researchers and the presence of intrinsic linguistic challenges, such as homonymy, the development of Deep Learning algorithms to address this issue has become widespread. Many AND deep learning methods have been developed, and surveys exist comparing the approaches in terms of techniques, complexity, performance. However, none explicitly addresses AND methods in the context of deep learning in the latest years (i.e. timeframe 2016-2024). In this paper, we provide a systematic review of state-of-the-art AND techniques based on deep learning, highlighting recent improvements, challenges, and open issues in the field. We find that DL methods have significantly impacted AND by enabling the integration of structured and unstructured data, and hybrid approaches effectively balance supervised and unsupervised learning.

\end{abstract}

\begin{keywords}
  Author name disambiguation \sep
  Machine Learning \sep
  Deep learning
\end{keywords}

\maketitle

\section{Introduction}

Author Name Disambiguation (AND) is a critical task for digital libraries where correctly associating publications with their respective authors is fundamental. This issue occurs when a set of publications contains ambiguous author names, e.g. when an author appears with different names (synonyms) in various papers or when several authors share the same name (homonyms) \cite{ferreira_brief_2012}. 
However, AND presents significant challenges, primarily due to the high number of authors sharing the same names and the presence of typographical errors in bibliographic metadata. Indeed, the implications of name ambiguity are extensive, impacting the reliability of, for instance, bibliometric analyses and the precision of citation-based metrics \cite{strotmann_author_2012}.

Reliable AND enables the creation of robust links between authors and their works, ensuring that research information about the authors is correctly attributed and enhancing the discoverability of academic output. Although identifiers such as ORCID could theoretically solve this problem, their limited adoption among researchers and the lack of their systematic use also in significant collections of bibliographic metadata (e.g. in OpenCitations \cite{peroni2020opencitations}) often obliges to rely on analysing bibliographic metadata to disambiguate author entities \cite{baglioni_reections_nodate}. 
These metadata can include the list of authors, their affiliations, the publication venue, and publication date/year, that are used in the context of approaches that leverage information coming from different bibliographic metadata repositories. 



AND can be distinguished between two sub-problems\cite{ferreira_brief_2012}: author assignment (AA) and author grouping (AG). The AA task links new publications to known author profiles, while the AG task clusters publications that belong to the same author. These sub-problems highlight the dual nature of AND, where the main challenge lies in managing both existing authors (usually identified with supervised techniques) and new, previously unregistered authors (usually detected with unsupervised techniques).
Name ambiguity is particularly evident in specific contexts, such as among authors with common surnames or working in regions where writing systems exploit different linguistic strategies (e.g. Chinese characters). The impact of these factors on machine learning approaches to AND has been widely demonstrated with emphases on the need for methods tailored to specific linguistic and cultural contexts \cite{10.1002/asi.24298}.


In the past, existing published reviews on AND provided taxonomies and valuable insights into traditional methods or graph-based approaches. For instance, Ferreira \emph{et al.} \cite{ferreira_brief_2012} propose a taxonomy for characterising the current AND methods described in the literature. The survey categorises automatic methods based on two main features: the evidence explored in the disambiguation task (web information, citation relationships, etc.) and the primary exploited approach type (author grouping and assignment methods). Sanyal \emph{et al.} \cite{sanyal_review_2021} focus on AND challenges in PubMed (\url{https://pubmed.ncbi.nlm.nih.gov}), the publication repository of the Life Science community. The survey proposes a taxonomy that classifies methods based on evidence explored (only co-authorship information or multiple metadata) and techniques used to generate similarity profiles (supervised, graph-based, and heuristic-based). De Bonis \emph{et al.} \cite{de_bonis_graph-based_2023}, instead, focus on graph-based approaches for AND. Their work comprehensively reviews popular graph-based methods, introduces a framework for comparing these approaches, and compiles a list of Scholarly Knowledge Graph (SKG) benchmarks used in the reviewed studies. They identify a taxonomy that classifies AND approaches based on three macro features: the learning strategy, the evidence explored, and the node representation strategy.

However, the above reviews do not focus on deep learning approaches within 2016-2024. The work introduced in this paper aims to present a literature review to fill this gap by systematically analysing deep learning methods applied to the AND task within the selected timeframe. It provides a comprehensive overview of the state-of-the-art methodologies, identifies their limitations, and highlights emerging trends. In particular, it focuses on how various approaches address metadata diversity across digital libraries, integrate structured and unstructured information, and improve performance in AA and AG tasks.


Within the scope of our systematic literature review, we aim to address the following research questions:
\begin{itemize}
\item How have deep learning approaches impacted AND resolution during recent years?
\item What limitations arise from deep learning AND algorithms?
\end{itemize}

The rest of this paper is organised as follows. Section~\ref{background} provides Background on the Author Name Disambiguation (AND) task, discussing key challenges, existing approaches, and gaps in the literature. Section~\ref{methodology} outlines the methodology adopted to search and select relevant literature and to classify the gathered data. Section~\ref{result} presents the results of our analysis where all the identified AND approaches have been categorised according to the particular deep learning technique adopted (supervised, unsupervised, and hybrid). Finally, Section~\ref{discussion} discusses key findings, focusing on dataset limitations and the impact of imbalanced data on AND models, and  Section~\ref{conclusions} concludes the paper and suggests future research directions.

\section{Background}\label{background}

The AND task is inherently challenging due to various factors, including name variations (e.g. abbreviations, translations, pseudonyms), transcription errors, and incomplete or incorrect metadata. For instance, the complexity of name polysemy highlights how personal names and organisations can appear in multiple forms—shortened, initialized, or abbreviated—leading to increased ambiguity \cite{Milojevic2013AccuracyOS}. In this context, we refer to name synonymity when we have two different structural representations of names indicating the same person (i.e. J.F. Kennedy and John Fitzgerald Kennedy) and to homonymous names when we have two identical names representing two different real people \cite{manzoor_toward_2022}.  
Similarly, differences in language and scripts may result in diverse representations of the same name (e.g., \begin{CJK*}{UTF8}{gbsn}毛澤東\end{CJK*} rendered as Mao Zedong or Mao Tse-tung), complicating automated retrieval and disambiguation processes \cite{Kim2023}. 

Another prominent issue is the scarcity of high-quality data. Errors in metadata, missing names, or pseudonyms create significant obstacles to constructing accurate predictive models, particularly for deep learning (DL) based approaches, which require large amounts of high-quality labelled data. While DL methods have shown potential in overcoming limitations of traditional or machine learning (ML) based approaches, thanks to their ability to integrate structured (e.g. author profiles) and unstructured (e.g. publication metadata) data and to model complex relationships, challenges remain. These include handling metadata heterogeneity across platforms and addressing imbalanced datasets, which often hinder the generalisability of algorithms.

Moreover, the lack of standard and reliable annotated datasets and shared resources for systematic evaluation limits the ability to compare methodologies and develop consolidated benchmarks directly.
One of the key challenges in AND research is establishing benchmarks to evaluate and compare performance with state-of-the-art methods. 
These datasets use unique author identifiers to link authors with their publications independently of metadata-provided names. By offering both the ground truth for the AND task (e.g. identifiers and associated names) and contextual information (e.g. co-authors), these datasets enable the development and evaluation of advanced disambiguation methods.

Despite the availability of labelled datasets for training supervised or semi-supervised methods, existing datasets have notable limitations. Many are manually created, which is time-intensive and prone to quality issues. Additionally, most datasets are small in scale or domain-specific (e.g. SCAD-zbMATH for mathematics), failing to capture the complexity of name ambiguities in large, real-world databases (Zhang, Li \& Lu, Wei \& Yang, Jinqing. (2021). Biases in datasets, such as overrepresentation of certain ethnicities or insufficient handling of name variations, further limit their applicability (Sanyal, D. K., Bhowmick, P. K., \& Das, P. P. (2021). These challenges underscore the need for robust datasets and methods to improve AND systems.

The methods analysed and revised in this survey take advantage of different benchmarks in the literature. Most of them rely on datasets derived from AMiner (https://www.AMiner.org), a free online service used to index, search, and mine big scientific data designated to identify connections between researchers, conferences, and publications. These benchmarks include:
\begin{itemize}
\item \textbf{AMiner-WhoIsWho}: the largest manually labeled AND benchmark \cite{Chen2023}. It features over 1,000,000 articles across three dataset versions, offering high-quality labels and extensive metadata. This dataset provides a robust foundation for evaluating state-of-the-art AND algorithms at scale.
\item \textbf{AMiner-534K} \cite{santini_knowledge_2022}: a knowledge graph extracted from an AMiner benchmark. Structural triples of the knowledge graph are split into training, testing, and validation for applying representation learning methods.
\end{itemize}

Other datasets used for testing methodologies to address the AND task are:
\begin{itemize}

\item \textbf{CiteSeerX} (https://citeseerx.ist.psu.edu), a digital library and search engine for scientific literature, primarily in computer and information sciences. It provides metadata for publications, including author names, abstracts, and citations, making it a valuable resource for developing and testing AND algorithms.
\item \textbf{DBLP} (https://dblp.org), a computer science bibliography that offers detailed metadata for millions of publications, including author names, titles, venues, and co-authorship networks. Its structured format and wide domain coverage make it a popular dataset for AND research, particularly for exploring co-author relationships.
\item \textbf{PubMed} (https://pubmed.ncbi.nlm.nih.gov) is a comprehensive database for life sciences and biomedical literature. It provides metadata such as author names, titles, abstracts, and affiliations. Its focus on biomedical research allows AND studies to address domain-specific challenges, including highly similar author names and complex collaboration networks.
\item \textbf{Scopus} (https://elsevier.com/products/scopus) is an indexing and abstract database resource with full-text links produced by Elsevier Co \cite{burnham_scopus_2006}.
\end{itemize}

A significant challenge in AND lies in the problems of over-merging and over-splitting. Over-splitting means that the same authors’ papers are incorrectly assigned into two or more clusters. Over-merging denotes that different authors' papers are assigned into one cluster. Existing AND methods used clustering or classification but ignored these two factors. 
As a result, over-splitting and over-merging problems will 
severely restrict the performance of a downstream algorithm 
tasks \cite{gong_more_2024}. 
Existing approaches for these tasks substantially rely on complex clustering-like architectures, and they usually assume the number of clusters is known beforehand or predict the number by applying another model, which involves increasingly complex and time-consuming architectures \cite{Zhang2020StrongBF}.

In summary, AND remains a dynamic and evolving area of research where integrating ML and DL approaches represents a significant step toward improving performance. Nevertheless, substantial challenges persist, including the need to handle heterogeneous data and address uncertainties related to the representation of real-world authors. These challenges highlight the importance of innovative and collaborative solutions to enhance the effectiveness of disambiguation techniques.

\section{Methodology}\label{methodology}

The literature review was conducted following a systematic and structured approach to ensure the inclusion of relevant, high-quality academic studies addressing AND. This process included defining inclusion and exclusion criteria, employing targeted search strategies, and constructing a comprehensive table to organize and analyze the selected resources.

\subsection{Research Preparation}
To guide the selection of academic works, specific inclusion criteria and exclusion criteria were defined. Studies were considered eligible for inclusion if they:

\begin{itemize}
\item Primarily addressed AND as a core component of their research. 
\item Were published between 2016 and 2024, ensuring the inclusion of recent advancements in the field.
\item Were published in peer-reviewed journals or conferences and made available in major bibliographic repositories.

\end{itemize}

These criteria ensured the review focused on impactful and high-quality contributions within the field.

\subsection{Literature Research Strategies}

The search for relevant studies was carried out using a combination of keyword-based and citation-based approaches:

\begin{itemize}

\item Google Scholar keyword search: The keywords "author name disambiguation" + "deep learning" were used to retrieve relevant studies. The first 50 results were selected, and their abstracts were reviewed to assess their alignment with the inclusion criteria.
\item PURE Suggest citation-based search: The PUREsuggest tool was employed to expand the pool of studies by leveraging citation-based recommendations. This approach incorporated snowballing techniques, where cited and citing articles were reviewed for relevance. The PURE suggestion process was repeated twice to ensure comprehensive coverage of key works in the field.

\end{itemize}

\subsection{Table Construction}

A structured table was created to analyse and compare the selected studies systematically. This table included the following columns, designed to capture essential details and provide a basis for qualitative evaluation:

\begin{itemize}
    
\item \textbf{Title}: Title of the resource.
\item \textbf{Abstract}: Abstract of the paper resource.
\item \textbf{Year}: Year of publishing.
\item  \textbf{DOI}: The publication year and DOI to ensure traceability.
\item  \textbf{Comparability}: Filter on resources comparability based on Datasets and Evaluation Metrics used. 
\item  \textbf{Type of Approach}: Whether the study addressed AND through assignment, grouping, or a combination of both.
\item  \textbf{Learning Strategy}: The classification of the methodological approach based on the learning paradigm adopted (whether supervised, unsupervised or using a mixed approach).
\item  \textbf{Dataset Selected}: Details about the dataset(s) used in the study.
\item  \textbf{DL Structure Summary}: A concise summary of the deep learning architecture implemented in the study.
\item  \textbf{Limitations}: Identified limitations or challenges discussed in the study.
\item  \textbf{Evaluation Metrics}: The metrics used to evaluate the performance of the proposed methodologies, such as accuracy, precision, recall, or F1 score.

\end{itemize}

This structured approach allowed for a systematic evaluation of the selected works, facilitating comparisons and highlighting trends, strengths, and gaps in the methodologies applied to AND.
The search result returned 52 documents, of which 28 were considered relevant after abstract scraping.

\section{Results and Analysis}\label{result}

The literature analysis within the selected timeframe highlights the dominant role of deep learning methodologies in author name disambiguation tasks. Specifically, the use of neural network-based models has become prevalent in addressing the challenges of disambiguation in complex academic networks. To better understand the diverse strategies employed, we categorize the various approaches based on their learning paradigm: supervised, unsupervised and some methods combining these approaches in hybrid setups. 
The rationale for this classification lies in the type of input data available in each study and whether the dataset contains disambiguated information about author entities and their publications. Supervised approaches typically rely on labelled datasets where author identities are explicitly annotated, enabling models to train with clear ground truth. Conversely, unsupervised methods operate in settings where such labelled data is unavailable, requiring clustering or other heuristic techniques to infer relationships. Given the difficulty of obtaining large-scale annotated datasets for training, hybrid systems have emerged as a promising solution. These methods aim to maximize the benefits of supervision while effectively leveraging available disambiguated data, often blending supervised learning with unsupervised strategies to enhance performance.
In the following sections, we provide a detailed examination of each category, discussing the methodologies employed, key contributions, and proposed solutions in the context of author name disambiguation.

\subsection{Supervised AND Techniques}

Deep learning supervised techniques make use of labelled training data manually created or collected from annotated databases, usually inputted into a classifier. In this setting, the inputs may for example include a set of publication features like co-authors, title words, year of publications and venue information, and the output is usually an author identifier/class.

\textbf{Capsule Neural Networks (CapsNets)}: Firdaus et al. (2021) \cite{9946586} proposed a CapsNet model for author classification using bibliographic data from DBLP. CapsNets encode author attributes, including name, co-authors, venue, title, and year, into activation vectors that capture entities' properties and spatial relationships. This enriched representation outperformed traditional neural networks, achieving an impressive 99.83\% accuracy on test data, highlighting the efficacy of CapsNets for AND.

\textbf{WhoIs Model}: Boukhers et al. (2022) \cite{boukhers_whois_2022} introduced the WhoIs model, a deep neural network designed for author disambiguation. The model uses Char2Vec embeddings to capture orthographic similarities between author names and BERT to generate contextualized embeddings for publication titles and sources. By training separate networks for each atomic name variant (e.g., initials and surname) and integrating co-author and research field information, WhoIs improves disambiguation accuracy.

\textbf{Bib2Auth Model}: Boukhers et al. (2021) \cite{boukhers_bib2auth_2021} also proposed Bib2Auth, a neural network model with two input layers. The first encodes the concatenation of co-author and content embeddings, where the content embedding is the average of title and source embeddings. The second input layer encodes the target author embedding. Author and co-author embeddings are generated using Char2Vec, which positions similar words closely in the representation space. Title and source embeddings are generated using BERT, providing contextualized word representations. Separating the inputs addresses content sparsity and enhances feature focus.

\textbf{Hybrid Deep Pairwise Classification}: Kim et al. (2019) \cite{kim_hybrid_2019} developed a hybrid approach combining structural and global features. Structural features, such as cosine similarities of attributes, are combined with global semantic features extracted via a DNN. Using Gradient Boosted Trees (GBT) for pairwise classification, the model determines whether two mentions belong to the same author, demonstrating the strength of hybrid frameworks in balancing feature complexity and classification efficiency.

\textbf{CONNA}: Zhao et al. (2022) \cite{chen_conna_2022} introduced CONNA, a framework designed for real-time AND as new articles are added to a system. It combines embedding techniques with pairwise classification to either assign publications to existing author profiles or create new ones. Notably, CONNA incorporates a "NIL" candidate for cases where an author is absent from the system and uses a multi-field profile (MFP) to capture attributes such as affiliations and publication venues.

\textbf{Co-Attention-Based Pairwise Learning}: Wang et al. (2023) \cite{wang_co-attention-based_2023} proposed a novel approach focusing on pairwise comparisons between author names using co-attention mechanisms. Their model integrates textual, categorical, and co-authorship attributes to capture the relationships between bibliographic features effectively. Self-attention and co-attention mechanisms ensure the model prioritizes the most relevant input features, enhancing disambiguation performance.

\textbf{Toward a New Paradigm for AND}: (Manzoor et al. 2022) \cite{manzoor_toward_2022} leveraged Convolutional Neural Networks (CNNs) for semantic name disambiguation using the PubMed dataset. The model processes sequences of words from attributes such as co-authors, affiliations, and publication titles, extracting meaningful semantic relationships. While the CNN architecture is not detailed, the study emphasizes the network's ability to capture patterns within bibliographic metadata.

\textbf{Deep Neural Network Structure to Improve Individual Performance based Author Classification}: 
Firdaus et al. (2019) \cite{firdaus_deep_2019} apply a DNN to a dataset of 125 Scopus publications from nine Indonesian authors, addressing both homonymy and synonymy. Data pairs were labeled (1 for the same author, 0 otherwise) and processed using Levenshtein distance with Z-score normalization. The best results (99.6\% accuracy) were achieved with 2–5 hidden layers using ReLU and Sigmoid activation functions.

\textbf{Identification of Indonesian Authors Using Deep Neural Networks}: (Firdaus et al., 2022) \cite{firdaus_identification_2022} developed a custom dataset consisting  of 125 publications  from  9 Indonesian  authors  with  the  following  attributes;  author  name,  title,  year,  source, author affiliation, co-authors name  instances extracted from Scopus to address AND for Indonesian authors. Preprocessing steps included normalization, label encoding, and PCA for dimensionality reduction. The DNN classifier incorporated ReLU activation functions in hidden layers and a softmax output layer. The result shows that the level of accuracy is 99.6\%on average.

\textbf{Strong Baselines for Author Name Disambiguation with and Without Neural Networks}: Zhang, Yu, Liu, \& Wang, 2020 \cite{Zhang2020StrongBF} This study introduces a robust pipeline combining pre-merging strategies (PMS) based on co-authorship, simple neural networks (SNN) for title-based semantic similarity, and post-blocking strategies (PBS) for cluster refinement. By integrating these techniques, the model efficiently clusters publications and improves disambiguation accuracy.

\textbf{Multiple Features Driven Author Name Disambiguation}:
Zhou, Chen, Wang, Xu,  Zhao, 2021 \cite{9590332} construct six similarity graphs (using the raw document and fusion feature) for each ambiguous author name. The structural information (global and local) extracted from these graphs is inputted into a novel encoder called R3JG, which integrates and reconstructs the information for an author. An author is therefore associated with four types of information: the raw document feature, the publication embedding based on the raw feature, the local structural information from the neighborhood, and the global structural information of the graph. Each node is embedded by using the random walk. The goal of the framework is to learn the latent information to enhance the generalization ability of the MFAND. Then, the integrated and reconstructed information is fed into a binary classification model for disambiguation.

\textbf{Towards Effective Author Name Disambiguation by Hybrid Attention}:
In Zhou et al., 2024 \cite{zhou_towards_2024} the Feature Extraction Model (EX) is composed of three hybrid attention mechanism layers, and each layer contains three key modules, i.e., a local structural perception, a global structural perception, and a feature extractor. Local structural perception consists of five hybrid GAT layers, whereas global structural perception consists of GCN and a single layer feedforward neural network. In the Decision Model (DI)  generated triplets are fed into an MLP classifier for disambiguation between each publication pair. 

\textbf{Pairwise Learning for Name Disambiguation in Large-Scale Heterogeneous Academic Networks}
Sun et al., 2020 \cite{sun_pairwise_2020} introduce Multi-view Attention-based Pairwise Recurrent Neural Network (MA-PairRNN) to solve the name disambiguation problem combining heterogeneous graph embedding learning and pairwise similarity learning into a framework. In addition to attribute and structure information, MA-PairRNN also exploits semantic information by meta-path and generates node representation in an inductive way. Furthermore, a semantic-level attention mechanism is adopted to fuse multiple meta-path based representations. A Pseudo-Siamese network consisting of two RNNs takes two paper sequences in publication time order as input and outputs their similarity.

\textbf{Name Disambiguation Scheme Based on Heterogeneous Academic Sites}:
Choi et al., 2023 \cite{choi_name_2024} propose a name disambiguation framework combining rule-based methods with deep learning. The rule-based component evaluates key attributes such as affiliation, publication year, and co-authors, assigning weights to attribute matches (e.g., exact title matches receive the highest weight). The deep learning component leverages a Graph Convolutional Network (GCN) to analyze vector representations of titles, keywords, and abstracts, effectively handling incomplete or inconsistent metadata.The GCN was performed using the created adjacency and feature matrices. Ultimately, based on the learned feature vectors, HAC was carried out for name disambiguation. A multi-classifier selects the most suitable approach based on available metadata, trained on labeled data and adaptable to future disambiguation schemes. 

\textbf{Author Name Disambiguation Using Multiple Graph Attention Networks}
Zhang et al. (2021) \cite{zhang_author_2021} explore the use of multiple Graph Attention Networks (GATs) for author name disambiguation. The academic network is represented as a heterogeneous graph, which is transformed into multiple homogeneous graphs based on meta-paths such as co-authorship and shared venues. Node features, including textual data from titles and abstracts, are processed with Word2Vec embeddings. GAT layers apply a multi-head attention mechanism to model the importance of neighboring nodes, and the resulting embeddings are fused using a fully connected layer. Clustering is performed using spectral clustering.

\textbf{Semantic Author Name Disambiguation with Word Embeddings}: 
Müller, MC. (2017) \cite{muller_data_2017} presents a supervised AND system that combines three models: co-author similarity, content similarity (using word embeddings for title matching), and metadata similarity. Each model is supervised, relying on labeled data for training. These models are integrated into a multi-layer neural network for final classification. For the clustering task, the system uses binary classifiers, adjusting the minimum positive confidence (mpc) threshold to control the precision and recall trade-off.

\subsection{Unsupervised AND Techniques}

Unsupervised approaches do not require labelled data and typically focus on clustering or embedding methods to disambiguate names. These methods are particularly useful when labelled datasets are unavailable or limited.

\textbf{BOND: Bootstrapping From-Scratch Name Disambiguation with Multi-task Promoting}:
The BOND method \cite{cheng_bond_2024} constructs multi-relational graphs to exploit higher-order information, jointly learning local and global representations within an end-to-end framework. This approach employs a Graph Attention Network (GAT) as an encoder to learn publication representations and integrates the DBSCAN clustering algorithm to optimize both local metric learning and global clustering within a unified structure.

\textbf{A Knowledge Graph Embeddings-based Approach for Author Name Disambiguation Using Literals} 
Santini et al. (2022) \cite{santini_knowledge_2022} propose a framework that leverages multimodal knowledge graph embeddings. The approach utilizes the LiteralE model to integrate literal information, such as publication titles and dates, into entity representations. Two variants are explored: LAND-glin, which incorporates textual embeddings derived from titles using a linear transformation, and LAND-ggru, which combines textual information and numeric literals via a Gated Recurrent Unit (GRU).

\textbf{Learning Semantic and Relationship Joint Embedding for Author Name Disambiguation}
Xiong et al. (2021) \cite{xiong_learning_2021} aim to bridge the gap between semantic and relational information in author name disambiguation. Their method encodes both semantic and relational information into a shared low-dimensional space, leveraging their complementarity and correlation to enhance performance. [...]

\textbf{Unsupervised Author Disambiguation Using Heterogeneous Graph Convolutional Network Embedding}
Qiao et al. (2019) \cite{qiao_unsupervised_2019} propose the use of Heterogeneous Graph Convolutional Networks (HGCN) to address author name disambiguation. Academic data are represented as a heterogeneous graph, where nodes correspond to publications and edges capture relationships such as co-authorship and shared venues. Initial node features are derived using Doc2Vec embeddings, while the HGCN employs relation-specific transformations and meta-path-guided random walks to learn embeddings that integrate structural and semantic information. Clustering is performed using a graph-enhanced hierarchical agglomerative clustering algorithm.

\textbf{Exploiting Higher Order Multi-dimensional Relationships with Self-attention for Author Name Disambiguation}
Pooja, Mondal, \& Chandra, 2022 \cite{pooja_exploiting_2022} propose an unsupervised approach based on graph convolutional networks (GCN) with a dual-level attention mechanism. The first level weighs the importance of relationships between documents (e.g., co-authorship or content), while the second level considers graph proximities. Multi-hop information is integrated to enhance node representations, improving the accuracy of author name disambiguation.

\subsection{Mixed AND Techniques}

Mixed AND approaches incorporate both supervised and unsupervised methods within their pipelines. Labeled data is used to train the supervised system, while these labels remain hidden from the unsupervised system. This type of approach leverages the availability of annotated data while addressing the challenge that such data is rarely available in large quantities.

\textbf{Name Disambiguation in AMiner: Clustering, Maintenance, and Human in the Loop}:
Zhang, Y., Zhang, F., Yao, P., \& Tang, J., 2018 \cite{zhang_name_2018} This framework tackles large-scale author name disambiguation by combining supervised and unsupervised methods. Supervised global metric learning uses a triplet-loss neural network to create unified document embeddings, while unsupervised local linkage learning refines these embeddings through a graph autoencoder. A supervised RNN estimates cluster sizes using pseudo-labeled data, and human feedback further enhances clustering accuracy. This hybrid approach balances labeled data usage with unsupervised refinement.

\textbf{MORE: Toward Improving Author Name Disambiguation in Academic Knowledge Graphs (AOG-BERT)}:
Gong, Fang, Peng, et al. (2024) \cite{gong_more_2024} propose the MORE framework that integrates multiple neural models, including OAG-BERT, SimCSE, LightGBM, and iHGAT, leveraging both supervised and unsupervised learning. The framework effectively combines semantic and structural features, achieving state-of-the-art performance for AND. The supervised components drive representation learning and cluster alignment, while the clustering process remains unsupervised.

\textbf{A Graph-Based Author Name Disambiguation Method and Analysis via Information Theory}:
Ma, Y., Wu, Y., \& Lu, C. (2020) \cite{ma_graph-based_2020} presents a hybrid approach for author name disambiguation, combining representation learning with both supervised and unsupervised techniques. Initially, word2vec is used for feature embedding, followed by a Graph Auto-Encoder (GAE) and Graph Embedding model for representation learning. The GAE and embedding model refine document representations based on document attributes and author relationships. Supervised learning fine-tunes the embeddings using labeled data, while unsupervised methods handle feature extraction and clustering. The framework integrates these methods to improve author disambiguation accuracy.

\textbf{Author Name Disambiguation via Heterogeneous Network Embedding from Structural and Semantic Perspectives}
Xie et al. (2022) \cite{xie_author_2022} introduce a method that considers both structural and semantic perspectives to embed publications into vectors. From the structural perspective, the approach uses weighted meta-path walks to sample context publications, capturing the heterogeneity of different meta-paths. Meta-path-level attention is employed to weigh various meta-paths and jointly learn overall structural representations. From the semantic perspective, Doc2Vec technology is applied to derive text embeddings of publications. These representations are combined and clustered using an adaptive clustering method.

\textbf{Bibliographic Name Disambiguation with Graph Convolutional Network}
Yan, Peng, Li, Li, \& Wang (2019)  \cite{cheng_bibliographic_2019} adopt an unsupervised model utilizing two customized GCNs: one for document embeddings (Document-GCN) and one for author embeddings (Person-GCN). Input graphs include relationships between authors (co-authorship), documents (keyword and metadata similarity), and bipartite connections between authors and documents. A triplet loss function ensures that embeddings bring related entities closer while pushing unrelated ones further apart. Clustering is performed using a hierarchical agglomerative clustering algorithm.

\textbf{Leveraging Knowledge Graph Embeddings to Disambiguate Author Names in Scientific Data}
Rettig et al., (2022) \cite{rettig_leveraging_2022} combines text and graph embeddings to enhance document representation and clustering. Named entity similarity derived from knowledge graphs is used as an additional source of information, integrating semantic insights from external sources such as knowledge graphs to improve disambiguation when co-authorship data is scarce.

\subsection{Method comparison}

The ideal way to compare state-of-the-art methods in deep learning is following these two criteria: 
\begin{itemize}

\item Studies must evaluate their results on the same or comparable datasets.
\item Studies must evaluate their results with the same evaluation metrics.

\end{itemize}

Most of the studies selected by this review rely on datasets derived from AMiner and evaluate their results with Precision, Recall and F1 metrics. For this reason we provide a detailed confrontation of them below. 

The results from the studies analyzed show significant variation in performance, with the highest F1-score achieved by Xie, Liu, Wang, and Jia (2022) \cite{xie_author_2022}, reaching an impressive 89.7, followed by Cheng, Chen, Zhang, and Tang (2024) \cite{cheng_bond_2024} with an F1-score of 87.72. Zhao et al. (2020) \cite{chen_conna_2022} achieved a score of 86.22 using the CONNA model with fine-tuning, while Xiong, Bao, and Wu (2021) \cite{xiong_learning_2021} achieved 84.7. Sun et al. (2020) \cite{sun_pairwise_2020} recorded an F1-score of 82.53, and Yan, Peng, Li, Li, and Wang (2020) \cite{yan_synergizing_2024} reached 81.3. The study by Gong, Fang, Peng, et al. (2024) \cite{gong_more_2024} showed an F1-score of 76.21, while Pooja, Mondal, and Chandra (2022) \cite{pooja_exploiting_2022} achieved 75.6.
At the lower end, Zhou, Chen, Wang, Xu, and Zhao (2021) \cite{9590332} obtained a score of 74.96, followed by Zhang et al. (2021) \cite{zhang_author_2021} with 73.13. Ma, Wu, and Lu (2020) \cite{ma_graph-based_2020} recorded 72.40, while Zhang, Yu, Liu, and Wang (2020) \cite{Zhang2020StrongBF} had a score of 70.19. Rettig, Baumann, Sigloch, and Cudré-Mauroux (2022) \cite{rettig_leveraging_2022} achieved 70.0, and Kim et al. (2019) \cite{kim_hybrid_2019} reached 69.14. Zhang, Y., Zhang, F., Yao, P., and Tang, J. (2018) \cite{zhang_name_2018} scored 67.79, while Santini et al. (2022) \cite{santini_knowledge_2022} had two models with F1-scores of 64.18 (LAND-glin) and 63.07 (LAND-ggru).
Table ~\ref{tab:disambiguation_methods} summarises the collected results for AMiner.


\begin{table*}
\caption{Summary of Author Name Disambiguation Methods evaluated on AMiner -- where ``S'' stands for supervised approach, ``U'' stands for unsupervised approach, and ``S+U'' stands for mixed approach.}
\label{tab:disambiguation_methods}
\resizebox{\textwidth}{!}{
\begin{tabular}{|m{4cm}|m{1.5cm}|m{3cm}|m{2cm}|m{1.5cm}|} 
\hline
\textbf{Article} & \textbf{Year} & \textbf{Dataset} & \textbf{Learning Strategy} & \textbf{F1-Score (AMiner)} \\
\hline
Xie, Liu, Wang, \& Jia (2022)\cite{xie_author_2022} & 2022 & AMiner + CiteSeerX & S+U & 89.7 \\
\hline
Cheng, Chen, Zhang, \& Tang (2024)\cite{cheng_bond_2024} & 2024 & AMiner-WhoIsWho v3 & U & 87.72 \\
\hline
Zhao et al. (2020)\cite{chen_conna_2022} & 2020 & AMiner & S & 86.22 \\
\hline
Xiong, Bao, \& Wu (2021)\cite{xiong_learning_2021} & 2020 & AMiner + DBLP + CiteSeerX & U & 84.7 \\
\hline
Sun et al. (2020)\cite{sun_pairwise_2020} & 2020 & AMiner + Semantic Scholar & S & 82.53 \\
\hline
Yan, Peng, Li, Li, \& Wang (2020)\cite{cheng_bibliographic_2019} & 2019 & AMiner & S+U & 81.3 \\
\hline
Qiao, Du, Fu, Wang, \& Zhou (2019)\cite{qiao_unsupervised_2019} & 2019 & AMiner + CiteSeerX & U & 78.60 \\
\hline
Gong, Fang, Peng, et al. (2024)\cite{gong_more_2024} & 2022 & AMiner-WhoIsWho + OAG & S+U & 76.21 \\
\hline
Pooja, Mondal, \& Chandra(2022)\cite{pooja_exploiting_2022} & 2022 & AMiner & U & 75.6 \\
\hline
Zhou, Chen, Wang, Xu, \& Zhao (2021)\cite{9590332} & 2021 & OAG-WhoIsWho + AMiner-AND & S & 74.96 \\
\hline
Zhang et al. (2021)\cite{zhang_author_2021} & 2021 & AMiner & S & 73.13 \\
\hline
Ma, Y., Wu, Y., \& Lu, C. (2020)\cite{ma_graph-based_2020} & 2020 & AMiner & S+U & 72.40 \\
\hline
Zhang, Yu, Liu, \& Wang (2020)\cite{Zhang2020StrongBF} & 2020 & AMiner & S+U & 70.19 \\
\hline
Rettig, Baumann, Sigloch, \& Cudré-Mauroux (2022)\cite{rettig_leveraging_2022} & 2022 & AMiner + SNSF & S+U & 70.0 \\
\hline
Kim et al. (2019)\cite{kim_hybrid_2019} & 2019 & AMiner + PubMed & S & 69.14 \\
\hline
Zhang, Y., Zhang, F., Yao, P., \& Tang, J. (2018)\cite{Zhang2020StrongBF} & 2018 & AMiner & S+U & 67.79 \\
\hline
Santini et al. (2022)\cite{santini_knowledge_2022} & 2022 & AMiner + OC & U & 64.18 \\
\hline
\end{tabular}
}
\end{table*}

Among those based on the DBLP dataset, we have \cite{9946586} with an F1 score of 0.98 and B\cite{boukhers_bib2auth_2021} with a score of 0.975, \cite{xiong_learning_2021} with a score of 0.875, and \cite{strotmann_author_2012} with a score of 0.737.

Among the methods based on the CiteSeerX dataset, we have \cite{xie_author_2022} with an F1 score of 71.9, \cite{boukhers_whois_2022} with a score of 71.3, \cite{qiao_unsupervised_2019} with a score of 69.8, and \cite{xiong_learning_2021} with a score of 68.1. 

For the methods based on the PubMed dataset, \cite{manzoor_toward_2022} achieves the highest F1 score of 95.95, followed by \cite{kim_hybrid_2019} with a score of 89.29.

Methods based on the Scopus dataset exploit \cite{firdaus_deep_2019} that achieves the highest F1 score of 99.6, followed by \cite{firdaus_identification_2022} with a score of 96.92.

For other datasets,\cite{wang_co-attention-based_2023} achieves an F1 score of 90.0 on the Dutch Central Catalogue (https://picarta.on.worldcat.org/discovery). Meanwhile, \cite{choi_name_2024}, which collects documents from various heterogeneous academic search services—including research papers, national R\&D reports, patents, and research reports—achieves an impressive score of 99.0.

\section{Discussion}\label{discussion}

Deep learning approaches have significantly advanced Author Name Disambiguation (AND) by utilizing neural network-based models to process complex academic data. These methods integrate both structured metadata (e.g., co-authorships, affiliations) and unstructured text (e.g., titles, abstracts), enabling the development of robust models for both author assignment and grouping tasks \cite{xie_author_2022,cheng_bond_2024,xiong_learning_2021,boukhers_whois_2022,zhang_author_2021,boukhers_bib2auth_2021}.
However, deep learning in AND has several limitations, including the lack of annotated datasets and data imbalance. In our review, the heavy reliance on AMiner as the primary dataset complicates the evaluation of these methods' generalizability, and this limitation also impacts the scalability of these models.
Overall, most studies analyzed in this review rely heavily on the AMiner dataset or its subsets. While this dataset is widely used, it comes with certain limitations. One notable issue is its selective concentration of Chinese names, which makes it particularly effective for disambiguating oriental names but less accurate when dealing with non-oriental names. Additionally, AMiner stands out for its reliability, as the data is automatically labelled and manually reviewed by human annotators. However, this high quality comes at the cost of significant human effort, limiting the availability of similar datasets for research.
Looking at ~\ref{tab:disambiguation_methods}, among the five top-performing models, the best result is achieved by a hybrid system, while the remaining approaches operate in either supervised or unsupervised settings.
The top three methods in our review demonstrate distinct methodologies in addressing AND, each suited to specific challenges in the task.
Xie, Liu, Wang, \& Jia (2022) \cite{xie_author_2022} employs a hybrid approach, integrating supervised and unsupervised techniques. It combines structural features, using meta-path-based attention, with semantic embeddings derived from textual data, creating a balanced method that excels in handling heterogeneous datasets with diverse data types. Cheng, Chen, Zhang, \& Tang (2024) \cite{cheng_bond_2024} takes an entirely unsupervised approach, focusing on scalability. By leveraging multi-relational graphs and clustering algorithms like DBSCAN, it avoids reliance on labeled data, making it particularly effective for large-scale datasets where annotations are scarce. Zhao et al. (2020) \cite{chen_conna_2022} , in contrast, adopts a supervised strategy, designed for real-time AND. Its strength lies in its dynamic capabilities, leveraging pairwise classification to assign publications to existing profiles or create new ones. A notable innovation is the introduction of a "NIL" candidate mechanism, allowing the system to handle previously unseen authors effectively.

Interestingly, despite the availability of AMiner's labeled data, the use of supervised settings does not appear to guarantee a significant boost in performance. This suggests that other factors, such as the method's design or the dataset's characteristics, may play a more critical role in achieving optimal results. Combining supervised and unsupervised deep learning approaches in a hybrid setup proves effective, as it retains the advantages of using labeled data for supervision while avoiding an increase in computational complexity \cite{gong_more_2024,boukhers_bib2auth_2021,zhang_name_2018}.

\section{Conclusions}\label{conclusions}
In this survey, we presented a comprehensive comparative study of deep learning-based AND methods developed between 2016 and 2024. Key challenges were highlighted, alongside significant advancements enabled by deep learning. These methods have become central to addressing the complexities of AND, leveraging structured metadata and unstructured text to achieve superior performance over traditional approaches. To reflect the diversity in learning strategies, we classified the approaches into supervised, unsupervised, and hybrid techniques. Notably, hybrid methods that combine supervised and unsupervised learning demonstrated state-of-the-art results by balancing the benefits of labeled data with scalability across diverse scenarios.

However, the study revealed a heavy reliance on AMiner as the primary dataset, which raises concerns about the generalizability of the evaluated methods. Data imbalances and the scarcity of annotated datasets further limit scalability, underscoring the pressing need for more diverse and standardized benchmarks.

In addressing the posed research questions, we conclude that deep learning has significantly advanced AND, particularly through the adoption of hybrid methods. Nonetheless, the lack of dataset diversity and standardized evaluation frameworks remains a critical challenge. Future work should prioritize the creation of comprehensive benchmarks and the development of richer datasets to ensure robust and generalizable solutions for academic name disambiguation.

\begin{acknowledgments}
  This work has been partially funded by the European Union’s Horizon Europe
framework programme under grant agreement No. 101095129 (GraspOS Project).
\end{acknowledgments}

\bibliography{litreview_biblio}
\end{document}